\documentclass[preprint,amsmath,titlepage,amssymb,english,aps,showkeys]{revtex4}
\usepackage{xcolor}
\usepackage{amssymb,amsmath}
\usepackage{graphicx}
\usepackage{hyperref}

\begin{document}

% Use the \preprint command to place your local institutional report
% number in the upper righthand corner of the title page in preprint mode.
% Multiple \preprint commands are allowed.
% Use the 'preprintnumbers' class option to override journal defaults
% to display numbers if necessary
%\preprint{}

%Title of paper

\title{A unimodular Kaluza-Klein theory}

% repeat the \author .. \affiliation  etc. as needed
% \email, \thanks, \homepage, \altaffiliation all apply to the current
% author. Explanatory text should go in the []'s, actual e-mail
% address or url should go in the {}'s for \email and \homepage.
% Please use the appropriate macro for each type of information

% \affiliation command applies to all authors since the last
% \affiliation command. The \affiliation command should follow the
% other information
% \affiliation can be followed by \email, \homepage, \thanks as well.
%\email[]{Your e-mail address}
%\homepage[]{Your web page}
%\thanks{}
%\altaffiliation{}

%\author{Richard Kerner}
%\affiliation{Universit\'e Sorbonne, France}

\author{J{\'u}lio C. Fabris}
\email{julio.fabris@cosmo-ufes.org}
\affiliation{N\'ucleo Cosmo-ufes\& Departamento de F{\'i}sica, CCE, Universidade Federal do Esp\'irito Santo, Vit\'oria, ES, Brazil}

\author{ Richard Kerner}
\email{richard.kerner@sorbonne-universite.fr}
\affiliation{Laboratoire de Physique Th\'eorique de la Mati\`ere Condens\'ee, Sorbonne-Universit\'e, Boite 121, 4 Place Jussieu, 75005, Paris,  France}
%Federal do Espirito Santo \\ Vitoria, ES, Brasil -- CEP: 29075-910.}
%
%
%\author{Richard Kerner}
%
%\affiliation
%\footnote{b: }
%
%

%Collaboration name if desired (requires use of superscriptaddress
%option in \documentclass). \noaffiliation is required (may also be
%used with the \author command).
%\collaboration can be followed by \email, \homepage, \thanks as well.
%\collaboration{}
%\noaffiliation

\date{\today}

% insert suggested PACS numbers in braces on next line
%\pacs{04.70.-s, 04.20.Dw, 98.80.Bp}
% insert suggested keywords - APS authors don't need to do this
%\keywords{}

%\maketitle must follow title, authors, abstract, \pacs, and \keywords

% body of paper here - Use proper section commands
% References should be done using the \cite, \ref, and \label commands

\begin{abstract}
Unimodular gravity became an object of increasing interest in the late $80$-ties (see, e.g. \cite{Henneaux}, \cite{Dragon}, \cite{Unruh})
and was recently used in primordial Universe modeling with cosmological constant, in the context of the Brans-Dicke
gravity including scalar field (\cite{Almeida}). In the present article we 
investigate the possibility of imposing the unimodular condition within the $5$-dimensional Kaluza-Klein theory including the
scalar field. The variational principle is formulated in $5$ dimensions first, and dimensional reduction is applied to
the resulting set of equations. A cosmological model based on these equations is then presented and discussed.
\end{abstract}


\maketitle

\section{Introduction}

\subsection{The Kaluza-Klein theory}

%\affiliation

%\affiliation

In the early $20^{\rm th}$ century Theodor Kaluza ($1885-1954$) {\cite{Kaluza}} and Oskar Klein ($1894-1977$) {\cite{Klein}} proposed 
a unified theory of gravity and electromagnetism, based on the Einsteinian General Relativity extended to five dimensions. By adding
an extra spatial coordinate $x^5$ and assuming that the pseudo-Riemannian geometry applies to the enlarged manifold,
Kaluza noticed that the $15$ independent components of metric tensor in $5$ dimensions can accomodate not only the $10$ components
of a $4$-dimensional subspace which can be identified with the metric tensor of General Relativity, but also the extra
four components of the $4$-potential of Maxwell's electromagnetism, if they are identified with mixed components ${\tilde{g}}_{5 \mu} = {\tilde{g}}_{\mu 5}$,
where $\mu, \nu = 0,1,2,3$
(In what follows, we shall denote by tilded symbols the geometric quantities defined on the $5$-dimensional Kaluza-Klein space, 
the corresponding $4$-dimensional entities such as the metric tensor, connection, curvature being denoted by the same symbols with no tilde upon them.):
\begin{equation}
{\tilde{g}}_{AB} = \begin{pmatrix} g_{\mu \nu} & A_{\mu} \cr A_{\nu} & 1 \end{pmatrix}
\label{gKK}
\end{equation} 
The standard variational principle applied to the Einstein-Hilbert lagrangian in $5$ dimension leads to $15$ partial differential equations;
therefore there is a risk that such a system is over-determined, given that we have only $14$ independent fields, $g_{\mu \nu}$
and $A_{\mu}$. 

However, by a happy coincidence, even with this incomplete version, the system was not over-determined due to the fact
that out of the $15$ Einstein equations in vacuo corresponding to the components of symmetric Ricci and metric tensors:
$(\mu \nu), \; \; (\mu 5 )$ and $(5 5)$ the last one ${\tilde{R}}_{55} - \frac{1}{2} {\tilde{g}}_{55} {\tilde{R}}$ reduces to tautology $0 = 0$, leaving exactly
$14$ equations, which are easily recognized as the usual $4$-dimensional Einstein's equations with electromagnetic energy-momentum
tensor as a source, along with Maxwell's equations coupled with gravitational field through covariant derivatives: the $15$ equations
\begin{equation}
{\tilde{R}}_{AB} - \frac{1}{2} {\tilde{g}}_{AB} {\tilde{R}} = 0, \; \; (A, B,..) = (\mu, 5), \; \; 
{\tilde{R}} = R - \frac{1}{4} F_{\lambda \rho} F^{\lambda \rho}.
\label{fiveeqs}
\end{equation}
with an extra assumption that the fields $g_{\mu \nu}$ and $A_{\mu}$ depend exclusively on space-time variables $x^{\mu}$ and not on $x^5$, 
when explicited give rise to the following system of coupled equations:
\begin{equation}
R_{\mu \nu} - \frac{1}{2}g_{\mu \nu} R = g^{\lambda \rho} F_{\mu \lambda} F_{\nu \rho} - \frac{1}{4} g_{\mu \nu} F_{\lambda \rho} F^{\lambda \rho},
\label{KKmunu}
\end{equation}
\begin{equation}
g^{\mu \nu} \nabla_{\mu} F_{\nu \lambda} = 0, 
\label{Maxwellcov}
\end{equation}
the last $55$ component reducing to $0 = 0$. This circumstance is often called ``the Kaluza-Klein miracle''.

In its first version proposed by Th. Kaluza, the fifth dimension was just an extra space coordinate, the
entire space being isomorphic with $M_4 \times R^1 \sim [ct, x, y, z, x^5]  \sim M_5$, a five-dimensional Minkowski space. 

A few years later, after the advent of Quantum Mechanics, Oskar Klein (\cite{Klein}) proposed to consider a compact fifth dimension, a circle with 
a very small radius such that it can not be detected by our current experiments (For example, it can be of the size of the Planck length). 
\vskip 0.5cm
\begin{figure}[hbt]
\centering
\includegraphics[width=4.4cm, height=2.5cm]{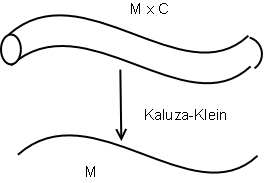} 
\label{fig:KKscheme}
{\caption{\small { The five-dimensional Kaluza-Klein space with a compact $5^{\rm th}$ dimension. The local coordinates are 
$x^A = (x^{\mu}, x^5); \; \; A = 1,2,..5, \; \; \mu, \nu,.. = 1,2,3,4 $,
which, under the projection $\pi$, reduce to points in the four dimensional space-time $M_4$:
$\pi (x^A) = \pi (x^{\mu}, x^5) = (x^\mu) \in M_4. $ }}}
\end{figure}

The dependence on the fifth dimension of functions defined on the  ``compactified'' space must be then periodic,
with a Fourier-like decomposition: 
\begin{equation}
f(x^{\mu}, x^5) = {\displaystyle{\sum_{k=0}^{\infty} }} a_k (x^{\mu}) e^{ik m x^5}.
\label{eik}
\end{equation}
 with dim[m] =cm$^{-1}$.

Then the eigenvalues of the fifth component of quantum momentum operator,  $p_5 = - i \hbar \partial_5$ 
are integer multiples of mass $m$, or perhaps the electric charge if one chooses an appropriate physical dimensional parametrisation.
Klein's aim was to explain the discrete values of electric charge of elementary particles known at that time, electrons and protons. 

In modern mathematical terms, such a structure is called a {\it principal fibre bundle}, 
denoted by $P(M, G)$, where $M$ denotes a differential manifold (in this case a pseudo-Riemannian space-time), 
and $G$ is a compact and semi-simple Lie group
(in this case the one-dimensional group $U(1)$, topologically equivalent to a circle).

The canonical projection $\pi: P(M, G) \rightarrow M$  maps points of $P(M, G)$ onto points in $M$, $\pi (p) = x \in M$.

The set of points in  $P(M, G)$ that project on the same point $x \in M$ is called {\it a fibre}, and is isomorphic with
 the structure group $G$ (here $U(1)$): $\pi^{-1} (x) \sim U(1).$ 
%\vskip 0.4cm
%\begin{figure}[hbt]
%\centering
%\includegraphics[width=4.4cm, height=2.5cm]{KK_scheme.png} 
%\label{fig:KKscheme}
%{\caption{\small { The five-dimensional Kaluza-Klein space. The local coordinates are 
%$x^A = (x^{\mu}, x^5); \; \; A = 1,2,..5, \; \; \mu, \nu,.. = (0, i) = 0,1,2,3, $
%which, under the projection $\pi$, reduce to points in the Minkowski space-time:
%$\pi (x^A) = \pi (x^{\mu}, x^5) = (x^\mu) \in M_4. $ }}}
%\end{figure}
%
%In its first version proposed by Th. Kaluza, the fifth dimension was just an extra space coordinate, the
%entire space being isomorphic with $M_4 \times R^1 \sim [ct, x, y, z, x^5]  \sim M_5$, a five-dimensional Minkowski space. 
% 
%Assuming that the compact fifth dimension is a circle with a very small radius, 
%the dependence on the fifth dimension of functions defined on the  ``compactified'' space must be periodic,
%with a Fourier-like decomposition: 
%\begin{equation}
%f(x^{\mu}, x^5) = {\displaystyle{\sum_{k=0}^{\infty} }} a_k (x^{\mu}) e^{ik m x^5}.
%\label{eik2}
%\end{equation}
% with dim (m) =$cm^{-1}$.
%
%Then the eigenvalues of the fifth component of quantum momentum operator,  $p_5 = - i \hbar \partial_5$ 
%are integer multiples of mass $m$ or their linear combinations, depending on the choice of solution. 

Later on, in the forties, the Kaluza-Klein theory was improved by P. Jordan \cite{Jordan} and Y. Thiry \cite{Thiry}
by inclusion of scalar field and its impact on generalized solutions. 

 Although the five-dimensional Kaluza-Klein theory is not viable, one of the reasons being the absence 
of chiral spinors in five space-time dimensions, it is present in some way or another in a more realistic theory 
as a restriction to the $U(1)$-subgroup of the full structural group. 
As such, it should describe fairly well the equations governing the electromagnetic sector
of any unified theory when other interactions (strong and electroweak ones) can be neglected.
 
\subsection{Enlarging the model by inclusion of scalar field}

Let us remind the full version of the Kaluza-Klein model, which englobes gravitational field given 
by the $4$-dimensional metric 
$g_{\mu \nu} (x)$, the electromagnetic field given by its $4$-potential $A_{\mu} (x)$ and the scalar field $\Phi (x)$, 
%$$ g_{\mu \nu} =  \eta_{\mu \nu} = {\rm diag} (+1, -1, -1, -1), \; \; \mu, \nu,.. = 0,1,2,3. $$
This ansatz corresponds to the $15$ degrees of freedom present in the $5$-dimensional Kaluza-Klein symmetric
metric tensor ${\hat{g}}_{AB}, \; \; 1,2, ... 5$.

The component $g_{55}$ of the five-dimensional metric tensor should be strictly negative in order to keep the fifth dimension spatial; 
this is why we shall give it the form $g_{55} = - \Phi^2 (x)$

Starting from this assumption, three particular situations can be studied separately now. We can consider a case with scalar field only being present, 
without the electromagnetic one.
This will lead to a variant of the tensor-scalar theory of gravitation, similar to the one proposed by Brans and Dicke (\cite{Brans}).

Another choice is the classical Kaluza-Klein model unifying gravitation and electromagnetism, but without scalar field. 
This choice amounts to suppressing one degree of freedom out of $15$, leaving only $14$ ones, 
the $4$-dimensional space-time metric $g_{\mu \nu}$
and the $4$-vector potential encoded in the components ${\tilde{g}}_{\mu 5} = {\tilde{g}}_{5 \mu}$ 
of the $5$-dimensional metric. 

Finally, we may consider the electromagnetic and scalar fields interacting in a flat Minkowskian space-time, 
in the absence of gravitation field considered as being negligible.

The five-dimensional metric with scalar field $\Phi(x)$ as the single degree of freedom remains diagonal:

\begin{equation}
{\tilde{g}}_{AB} = {\rm diag} \left(+1, -1, -1, -1, -\Phi^2(x) \right).
\label{gABdiag}
\end{equation} 

In principle, the notation $\Phi(x)$ can mean the dependence of the scalar field $\Phi$ not only on the space-time coordinates
$ (x^0=ct, x^1, x^2, x^3) $ but also on the fifth coordinate $x^5$, so that in principle we may have 
not only $\partial_{\mu} \Phi \neq 0$, but also $\partial_5 \Phi \neq 0$. 
It is also easily seen that the determinant of ${\tilde{g}}_{AB}$ is equal to $ \Phi^2$.

However, supposing that the fifth dimension is the structural group $U(1)$, i.e. is homeomorphic to a circle $S^1$, the dependence
of $\Phi$ on $x^5$ can be only a periodic one:
\begin{equation}
\Phi ( x^{\mu}, x^5 ) = \cos (\tilde n \; e \; x^5 + \delta ) \cdot \Phi (x^{\mu}), \; \; {\rm so \; that} \; \; \partial_{5}^2 \Phi = - \tilde n^2 e^2 \Phi,
\label{partial55}
\end{equation}  
where $\tilde n$ correspond to the normal mode.

Let us derive the set of general formulas for metrics, connections and curvature in $5$ dimensions, 
with all the $15$ degrees of freedom present.

The calculus in coordinates turns out to be quite complicated, but introducing the non-holonomic local frames 
simplifies the computations considerably.

%The basis of $1$-forms is chosen to be:
%
%\begin{equation}
%\theta^{\mu} = d x^{\mu}, \; \; \; \; \theta^5 = dx^5 + k \; A_{\mu} dx^{\mu},
%\label{thetas}
%\end{equation}
%\textcolor{blue}{to define $k$?}
%

The non-holonomic local frame is defined by the following choice of local basis of $1$-forms:
\begin{equation}
\theta^{\mu} = d x^{\mu}, \; \; \; \; \theta^5 = dx^5 + k \; A_{\mu} dx^{\mu},
\label{thetas}
\end{equation}
the constant $k$ must have the dimension of length ($cm$) in order to ensure the uniformity with space variables $x^{\mu}$.  
%The $1$-forms:
%\begin{equation}
%{\rm The \; 1-forms:} \; \; \; \theta^{\mu} = d x^{\mu}, \; \; \; \; \theta^5 = dx^5 + k \; A_{\mu} dx^{\mu},
%\label{thetas}
%\end{equation}
The dual vector fields satisfies $\theta^A ({\cal{D}}_B) = \delta^A_B$:
\begin{equation}
{\cal{D}}_{\mu} = \partial_{\mu} - k \; A_{\mu} \partial_5, \; \; \; \; {\cal{D}}_5 = \partial_5.
\label{CalD}
\end{equation}
Introducing transition matrices $U^A_B$ and ${\overset{-1}{U^B_C}}$ such that $\theta^A = U^A_B dx^B, \; {\cal{D}}_C = {\overset{-1}{U^D_C}} \partial_D $
we can write:
$$U^{\mu}_{\nu} = \delta^{\mu}_{\nu}, \; \; \; U^{\mu}_5 = 0, \; \; U^5_{\mu} = k A_{\mu}, \; \; U^5_5 = 1;$$
\begin{equation}
{\overset{-1}{U^{\mu}_{\nu}}} = \delta^{\mu}_{\nu}, \; \; {\overset{-1}{U^{5}_{\nu}}} = - k A_{\nu}, \; \; {\overset{-1}{U^{\mu}_{5}}} = 0 
\; \;  {\overset{-1}{U^{5}_{5}}} = 1.
\label{UUinv}  
\end{equation}
Quite obviously, both determinants are equal to $1$: det(($U^A_B) = 1$ and det (${\overset{-1}{U^B_C}}) = 1.$ 
The metric tensor expressed in the non-holonomic frame is easily deduced from the square of the  $5$-dimensional length element,
taking on the following form:
\begin{equation}
ds^2 = g_{\mu \nu} dx^{\mu} dx^{\nu} - \Phi^2 \left[ dx^5 + k \; A_{\mu} dx^{\mu} \right] \left[ dx^5 + k \; A_{\nu} dx^{\nu} \right]
\label{deessquare}
\end{equation}
leading to the following $5 \times 5$ matrix representation:
\begin{equation}
{\tilde{g}}^{AB} = \begin{pmatrix} g_{\mu \nu} + k^2 \Phi^2 A_{\mu} A_{\nu} & - k \Phi^2 A_{\nu} \cr - k \Phi^2 A_{\mu} & - \Phi^2 \end{pmatrix} 
\label{gABfive}
\end{equation}
The inverse matrix becomes then:
\begin{equation}
{\tilde{g}}^{BC} = \begin{pmatrix} g^{\nu \lambda}  &  k A_{\lambda} \cr k  A_{\nu} & - \Phi^{-2} + k^2 A^{\nu} A^{\lambda} \end{pmatrix} 
\label{invgABfive}
\end{equation}
One easily checks that 
$${\tilde{g}}_{AB} {\tilde{g}}^{BC} = \delta^A_C.$$
It is also obvious that due to the fact that the transition matrix $U^A_B$ is unimodular, the determinant of the metric ${\tilde{g}}_{AB}$
is the same as that of the metric $g_{AB}$, i.e. equal to $\Phi^2$. 

The explicit calculus of the connection and the Riemann tensor components in the non-holonomic frame
can be found elsewhere (e.g. R. Kerner, 1981), and we give them in the Appendix I at the end of the article. Here we need only the final expression
for ${\tilde{R}}_{AB}$ and ${\tilde{R}}$ in order to formulate Einstein's equations in $5$ dimensions.

The resulting expression for the $5$-dimensional curvature is quite simple indeed:
\begin{equation}
{\tilde{R}} = {\overset{4}{R}} - \frac{1}{4} \Phi^2 \; F_{\mu \nu} F^{\mu \nu} - \frac{2}{3 \Phi^2} g^{\mu \nu} \partial_{\mu} \Phi \partial_{\nu} \Phi.
\label{Rexpl}
\end{equation}
Considered as the integrand of a $5$-dimensional variational principle, this lagrangian density will lead to the following Einstein's equations
when varying with respect to the metric only:
\begin{equation} 
{\tilde{R}}_{AB} - \frac{1}{2} {\hat{g}}_{AB} {\tilde{R}} = 8 \pi G \left[ T_{AB}^{(\Phi)} + \frac{k^2}{16 \pi G} T_{AB}^{(F)} \right]
\end{equation}
where formally
\begin{equation}
T_{AB}^{(\Phi)} = \partial_A \Phi \partial_B \Phi - \frac{1}{2} {\tilde{g}}_{AB} ({\tilde{g}}^{CD} \partial_C \Phi \partial_D \Phi ),
\label{TPhi}
\end{equation}
and 
\begin{equation}
T_{AB}^{(F)} = F_{AC} F^C_{\; \; B} - \frac{1}{4} {\tilde{g}}_{AB} (F_{CD} F^{CD} ).
\label{TPeF}
\end{equation}
When the dependence $\Phi$ on $x^5$ of the $\tilde n-th$ mode is periodic, only the external space-time components are
different from zero:
\begin{equation}
T_{\mu \nu}^{(\Phi)} = \partial_{\mu} \Phi \partial_{\nu} \Phi - \frac{1}{2} {\hat{g}}_{\mu \nu} \left[ {\hat{g}}^{\lambda \rho} 
 \partial_{\lambda} \Phi \partial_{\rho} \Phi - n^2 e^2 \Phi^2 \right] ,
\label{TPhimunu}
\end{equation}
(where we neglected the mixed terms with $F_{\mu \nu}$ ) and 
\begin{equation}
T_{\mu \nu}^{(F)} = F_{\mu \lambda} F^{\lambda}_{\; \; \nu} - \frac{1}{4} {\hat{g}}_{\mu \nu} (F_{\lambda \rho} F^{\lambda \rho} ).
\label{TPeFmunu}
\end{equation}

The variation with respect to the scalar field $\Phi$ and the $4$-vector potential $A_{\mu}$ leads to the following equations of motion:
\begin{equation}
\frac{1}{\Phi} \partial_{\mu} \left[ \Phi \; F^{\mu \nu} \right] = 0,
\label{MotionF}
\end{equation}
and 
\begin{equation}
(\box \Phi + \tilde n^2 e^2 ) \Phi = 0.
\label{boxPhi} 
\end{equation}
where the term $\tilde n^2 e^2$ comes from the second derivative of $\Phi$ with respect to the circular coordinate $x^5$ 
and plays the role of a mass term for the Klein-Gordon scalar field equation. As it was underlined before, 
the presence of the $4$-vector potential
does not influence the unimodular condition, because $A_{\mu}$ does not contribute to the determinant of the $5$-dimensional metric. 

\section{Cosmological model in $5$ dimensions}
\label{blue}

\subsection{The generalized FRW metric} 

Initially, the Kaluza-Klein model was intended to incorporate Maxwellian electromagnetism into Einstein's General Relativity. The first cosmological applications
can be traced down to the Brans-Dicke scalar-tensor theory which turned out to be isomorphic with a variant of Kaluza-Klein theory including a scalar field
as the $(55)$ component of metric tensor. 

In $1980$ Chodos and Detweiler \cite{CD} proposed a vacuum Kasner-type cosmological solution in the $5$-dimensional Kaluza-Klein space.
The metric element for this model was
\begin{equation}
ds^2 =dt^2 - {\sqrt{t}} \left[ dx^2 + dy^2 + dz^2 \right] - \frac{1}{\sqrt{t}} \rho^2 d \chi^2.
\label{cosmods}
\end{equation}
where the last angular variable $\chi$ comes from the fifth cyclic dimension, and $\rho$ is its radius. 
%\textcolor{blue}{$\rho$ seems not defined.}

This metric can be generalized to more extra dimensions ($n$ to be more precise):

\begin{equation}
ds^2 = dt^2 - {\displaystyle{\sum_{i=1}^3}} t^{2 k_i} (dx^i)^2 - {\displaystyle{\sum_{a=4}^{3+n}}} t^{2 k_a} (dy^a)^2
\label{Kasnercosm}
\end{equation}
satisfying the following conditions:
\begin{equation}
{\displaystyle{\sum_{i=1}^3}} k_i + {\displaystyle{\sum_{a=4}^{3+n}}} k_a = 1, 
\end{equation}
\begin{equation}
{\displaystyle{\sum_{i=1}^3}} k_i^2 + {\displaystyle{\sum_{b=4}^{3+n}}} k_b^2 = 1
\end{equation}

The Friedmann-Robertson-Walker metric can be naturally generalized if we assume that the extra space dimensions 
form a compact spherically symmetric manifold. Then the overall metric can be derived from the following line element squared:
\begin{equation}
ds^2 = dt^2 - R_d^2 (t) \; g_{ij } dx^i dx^j - R_n^2 (t) g_{ab} dy^a dy^b,
\label{FRWplus}
\end{equation}
with two time-dependent scale factors, $R_d (t)$ for the space dimensions of our space-time, $d=3$, and $R_n (t)$ for the  
internal $n$-dimensional compact space - most usually, a $n$-dimensional sphere. This ansatz yields the following Ricci tensor:
$$R_{00} = - \left[ 3 \frac{{\ddot{R}}_d}{R_d} + n \frac{{\ddot{R}}_n}{R_n} \right],$$
\begin{equation} 
R_{ij} = \left[ \frac{2k_d}{R_d^2} + \frac{d}{dt} \left( \frac{{\dot{R}_d}}{R_d} \right) + \frac{{\dot{R_d}}}{R_d} 
\left(3 \frac{{\dot{R_d}}}{R_d} + n \frac{{\dot{R_n}}}{R_n}\right) \right] g_{ij}, 
\label{ThreeR}
\end{equation}
$$ R_{ab} = \left[ \frac{2k_n}{R_n^2} + \frac{d}{dt} \left( \frac{{\dot{R}_n}}{R_n} \right) + \frac{{\dot{R_n}}}{R_n} 
\left(3 \frac{{\dot{R_d}}}{R_d} + n \frac{{\dot{R_n}}}{R_n}\right) \right] g_{ab}, $$

In $1985$ D. Sahdev \cite{Sahdev} obtained solutions of this system with several perfect fluids added on the right-hand side. The nice feature
was that $R_d$ was increasing with time, and $R_n$ decreasing. However, instead of stabilizing at some small but finite value,
as physics would require, the internal radius $R_n$ tended to zero. 

\subsection{The $f(R)$ and higher order lagrangians}

In all models mentioned above the field equations were derived exclusively from the Einstein-Hilbert variational principle 
involving Riemannian scalar curvature as the unique (up to a constant cosmological term eventually) integrand. Possible 
inclusion of higher-order invariants of the Riemann tensor, in particular the quadratic Gauss-Bonnet term in the context of
the Kaluza-Klein theory and its non-abelian generalizations (\cite{Kerner1981}), was almost systematically ignored, in spite
of the throughout discussions of the Gauss-Bonnet lagrangians by Lanczos (\cite{Lanczos}) and Lovelock (\cite{Lovelock}).
The first attempt to include the Gauss-Bonnet term in the lagrangian of the $5$-dimensional Kaluza-Klein model was
made in $1987$ in order to produce a simple model of non-linear electrodynamics (see {\cite{Kerner1987}) 

On the other hand, various non-linear generalizations of General Relativity based on lagrangians using a function
of scalar curvature, $f(R)$, have gained certain popularity since the early $80$-ties \cite{Star,Kerner1982,Duruisseau,Odintsov,Capozziello}. 
The main problem haunting the $f(R)$ cosmological models was the
fact that they led to the fourth-order differential equations, reducible to the third-order ones by means of the Bianchi
identities, but no lower than third order. The Gauss-Bonnet invariant is the only quadratic combination leading to the 
second-order differential equations. The cubic and quartic Gauss-Bonnet invariants display the same poperty.

A cosmological model using the Gauss-Bonnet invariants in the generalized Kaluza-Klein space was introduced in $1988$ 
by B. Giorgini and R. Kerner (\cite{Giorgini1988}). It was realized as a $10$-dimensional
double fibre bundle $P( P(V_4, SU(2)), SU(2))$ with two structural $SU(2)$ groups interpreted as ``internal'' spaces:
and the $4$-dimensional space-time. The corresponding $10$-dimensional FRW metric contains three scale factors:
\begin{equation}
{\tilde{g}}_{AB} = {\rm{diag}} (c^2 dt^2 - a^2 (t) (dr^2 + r^2 d \Omega) - b^2(t) \; d^3 \xi - c^2 (t) \; d^3 \chi).
\label{tendims}
\end{equation}
where $d \Omega$ stands for the solid angle $d \theta^2 + \sin^2 \theta d \Phi^2$, while $d^ \xi$ and $d^ \chi$
 are invariant volume elements defined independently on the two $SU(2)$ structural groups. The scale factor $a(t)$ describes
the cosmic time behavior of the four-dimensional space, while the factors $b(t)$ and $c(t)$ pertain to the time-dependent scales
of two ``internal'' spaces. 

Finding explicit analytical solutions was obviously too difficult for such a highly non-linear (up to the sixth order) system; 
nevertheless the singular points in the phase space of solutions were found, and the qualitative behavior of the three scale factors
could be determined. While one of the internal spaces was growing and the other one shrinking, the four-dimensional space time
was following an exponential expansion, fed by the energy released by the internal spaces energy exchange.

\section{ Kaluza-Klein unimodular gravity through a variational principle}

Unimodular gravity was first proposed by Einstein in 1919 \cite{ein} in order to fix a coordinate system
by imposing a condition on the determinant of the metric, coinciding with that of the Minkowski space-time in cartesian coordinates, 
that is, $\sqrt{-g} = 1$. This proposal has evolved more recently and the unimodular gravity was viewed as another approach 
to the cosmological constant problem and to quantization of gravity, see for example \cite{Unruh,wein,brito,bran,u1}. 
In fact, the unimodular condition implies that the resulting field equations, after solving the constraint on the determinant of the metric, are traceless. Due to this property, any cosmological constant term disappears from the field equations. However, it can reappear as an integration constant. One consequence of this property is that, for example, the possible connection of the cosmological constant, for example, with the vacuum energy is lost. Notice, moreover, that instead of imposing the determinant of the metric equal to 1, the unimodular condition can be generalized by imposing that the determinant of the metric is equal to the determinant of a given fiducial metric. In any case, the condition on the determinant implies that the theory is invariant by a restricted class of diffeomorphism, the transverse diffeomorphism. 
By generalizing the unimodular gravity to the Kaluza-Klein, we follow the lines sketched in \cite{bran} (see also \cite{u2}).

Our investigation of unimodular gravity in the context of the $5$-dimensional Kaluza-Klein theory should be
preceded by the following considerations. We are confronted with an alternative choice of the way we apply the unimodularity
condition. We can choose to impose it to the variational principle in $5$ dimensions, find the solutions and then proceed
to dimensional reduction, or perform dimensional reduction first, expressing everything in terms of $4$-dimensonal Riemannian
metric and electromagnetic and scalar fields, and then proceed to variational principle in $4$ dimensions with unimodular
condition imposed on the $4$-dimensional metric. The situation is illustrated by the following diagram:
  
\begin{figure}
\centering
\includegraphics[width=8.8cm, height=4.5cm]{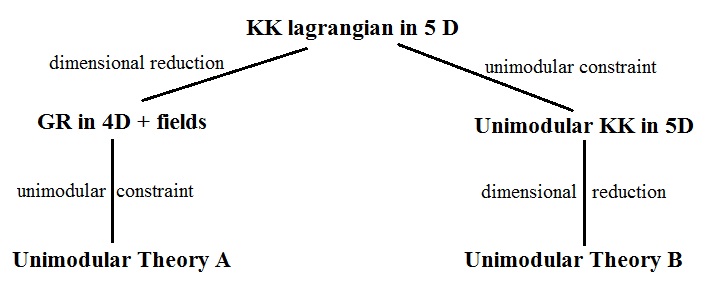} 
\label{KKunimod}
\caption{Possible dimensional compactification of the Kaluza-Klein theory leading to different effective theories in four dimensions.}
\end{figure}

In order to explore all the degrees of freedom present in the Kaluza-Klein setting, we choose to apply the unimodular condition
directly to the $5$-dimensional variational principle.

Our starting point is the following lagrangian density:
\begin{eqnarray}
{\cal L} = \sqrt{\tilde{g}}{\tilde{R}} + \lambda(\sqrt{\tilde{g}} - \xi) + \sqrt{\tilde{g}} {\cal L}_m.
\end{eqnarray}
The tildes indicate the five dimensional quantities; $\lambda$ is the lagrangian multiplier fixing the unimodular condition.

Using the variational principle:
\begin{eqnarray}
\label{eq1}
\tilde R_{AB} - \frac{1}{2}\tilde g_{AB}\tilde R - \frac{\lambda}{2}{\tilde{g}}_{AB} &=& 8\pi G T_{AB}, \\
\label{eq2}
\sqrt{\tilde g} &=& \xi.
\end{eqnarray}
The external function (not subject to the variational principle) $\xi$ allows to fix the determinant of the volume 
in a given coordinate system.

Taking the trace of (\ref{eq1}), it results in the following relation:
\begin{eqnarray}
\lambda = \frac{2 - D}{D}\tilde R - \frac{16\pi G}{D} T,
\end{eqnarray}
where $D = 4 + n$, $n$ being the number of extra-dimension.

Inserting this relation again into (\ref{eq1}), we obtain:
\begin{eqnarray}
\label{eq-u1}
\tilde R_{AB} - \frac{1}{D}\tilde g_{AB}\tilde R &=& 8\pi G \biggr(\tilde T_{AB} - \frac{1}{D}\tilde g_{AB}\tilde T\biggl), \\
\sqrt{\tilde g} &=& \xi.
\end{eqnarray}

Using the Bianchi identities in (\ref{eq-u1}), we obtain,
\begin{eqnarray}
\frac{D - 2}{2D}\tilde R^{;A} = 8\pi G\biggr\{{\tilde T^{AB}}_{;B} - \frac{\tilde T^{;A}}{D}\biggl\}.
\end{eqnarray}
If the the conservation of the energy-momentum tensor is imposed, ${\tilde T^{AB}}_{;B} = 0$, it is possible to write,
\begin{eqnarray}
\frac{D - 2}{2D}\tilde R + 8\pi G\frac{\tilde T}{D} = - \tilde\Lambda,
\end{eqnarray}
where $\tilde\Lambda$ is an integration constant which can be identified with the multidimensional cosmological constant.
The field equations become then,
\begin{eqnarray}
\label{eq-u2}
\tilde R_{AB} - \frac{1}{2}\tilde g_{AB}\tilde R = 8\pi G \tilde T_{AB} + g_{AB}\tilde\Lambda.
\end{eqnarray}
The imposition of the conservation of the energy-momentum tensor is, however, an extra hypothesis.

In order to preserve the unimodular condition (\ref{eq2}) under an infinitesimal coordinate transformation, 
\begin{eqnarray}
x^A \quad \rightarrow \quad x^A + \epsilon^A,
\end{eqnarray}
the transversal condition
\begin{eqnarray}
\bar\nabla_A\epsilon^A = 0.
\end{eqnarray}
The bar indicates that the covariant derivative is defined in terms of the fiducial metric related to the unimodular condition, see Ref. \cite{u1}.
This restriction generalizes, in $D$ dimension, the corresponding condition in four dimension: the unimodular theory in higher dimensions 
is still invariant by the transversal class of diffeomorphism (for a discussion on transversal diffeomorphism, see for example \cite{td}).

\section{Reduction to four dimensions: the cosmological setup}

Let us proceed by constructing a four dimensional theory out of the multidimensional theory. We change slightly the notation used in section \ref{blue}.
We write the multidimensional metric, with dimension $D = 4 + n$, as follows:
\begin{eqnarray}
ds^2 = g_{\mu\nu}dx^\mu dx^\nu - \Phi^2\delta_{ab}dx^a dx^b. 
\end{eqnarray}
The nonvanishing components of the Christoffel symbols
\begin{eqnarray}
\tilde\Gamma^C_{AB} = \frac{1}{2} {\tilde{g}}^{CD} \biggr(\partial_A \tilde g_{DB} + \partial_B\tilde g_{DA} - \partial_D\tilde g_{AB}\biggl),
\end{eqnarray}
are:
\begin{eqnarray}
\tilde\Gamma^\rho_{\mu\nu} &=& \Gamma^\rho_{\mu\nu},\\
\tilde\Gamma^{\rho}_{ab} &=& \Phi\Phi^{;\rho}\delta_{ab},\\
\tilde\Gamma^a_{\rho b} &=& \frac{\Phi_{;\rho}}{\Phi}\delta_{ab}.
\end{eqnarray}

The nonvanishing components of the Ricci tensor are:
\begin{eqnarray}
\tilde R_{\mu\nu} &=& R_{\mu\nu} - n\frac{\Phi_{;\mu;\nu}}{\Phi}, \\
\tilde R_{ab} &=& \biggr\{\Phi\Box\Phi + (n - 1)\Phi_{;\rho}\Phi^{;\rho}\biggl\}\delta_{ab}.
\end{eqnarray}
The Ricci scalar is:
\begin{eqnarray}
\tilde R = R - 2n\frac{\Box\Phi}{\Phi} - n(n - 1)\frac{\Phi_{;\rho}\Phi^{;\rho}}{\Phi^2}.
\end{eqnarray}

The components of the gravitational tensor
\begin{eqnarray}
\tilde G_{AB} = \tilde R_{AB} - \frac{1}{2}\tilde g_{AB}\tilde R,
\end{eqnarray}
are:
\begin{eqnarray}
\tilde G_{\mu\nu} &=& G_{\mu\nu} - n \biggr\{\frac{\Phi_{;\mu;\nu}}{\Phi} - g_{\mu\nu}\frac{\Box\Phi}{\Phi}\biggl\} 
+ \frac{n(n - 1)}{2}g_{\mu\nu}\frac{\Phi_{;\rho}\Phi^{;\rho}}{\Phi^2},\\
\tilde G_{ab} &=& \frac{\Phi^2}{2}\biggr\{R - 2(n - 1)\frac{\Box\Phi}{\Phi} 
 - (n - 1)(n - 2)\frac{\Phi_{;\rho}\Phi^{;\rho}}{\Phi^2}\biggl\}\delta_{ab}.
\end{eqnarray}

The components of the unimodular gravitational tensor
\begin{eqnarray}
\tilde E_{AB} = \tilde R_{AB} - \frac{1}{D}\tilde g_{AB}\tilde R,
\end{eqnarray}
are,
\begin{eqnarray}
\tilde E_{\mu\nu} &=& R_{\mu\nu} - \frac{1}{n+ 4}g_{\mu\nu}R - n\frac{\Phi_{;\mu;\nu}}{\Phi} + \frac{1}{n + 4}g_{\mu\nu}\biggr\{2n\frac{\Box\Phi}{\Phi}
 + n(n - 1)\frac{\Phi_{;\rho}\Phi^{;\rho}}{\Phi^2}\biggl\},\\
\tilde E_{ab} &=&  \frac{\Phi^2}{n + 4}\biggr\{R - (n - 4)\frac{\Box\Phi}{\Phi} + 4(n - 1)\frac{\Phi_{;\rho}\Phi^{;\rho}}{\Phi^2}\biggl\}\delta_{ab}.
\end{eqnarray}

The specificity of the construction displayed here can be verified already in the vacuum case. With no matter component, the unimodular field equations read,
\begin{eqnarray}
R_{\mu\nu} - \frac{1}{n+ 4}g_{\mu\nu}R = n\frac{\Phi_{;\mu;\nu}}{\Phi} - \frac{1}{n + 4}g_{\mu\nu}\biggr\{2n\frac{\Box\Phi}{\Phi}
 + n(n - 1)\frac{\Phi_{;\rho}\Phi^{;\rho}}{\Phi^2}\biggl\},\\
R = (n - 4)\frac{\Box\Phi}{\Phi} - 4(n - 1)\frac{\Phi_{;\rho}\Phi^{;\rho}}{\Phi^2}.
\end{eqnarray}
When $n = 1$, the equations become:
\begin{eqnarray}
R_{\mu\nu} - \frac{1}{5}g_{\mu\nu}R = \frac{\Phi_{;\mu;\nu}}{\Phi} - \frac{2}{5}g_{\mu\nu}\frac{\Box\Phi}{\Phi},
\\
R = - 3 \frac{\Box\Phi}{\Phi}.
\end{eqnarray}
These equations can be rewritten as,
\begin{eqnarray}\label{TB1}
R_{\mu\nu} &=& \frac{1}{\Phi}(\Phi_{;\mu;\nu} - g_{\mu\nu}\Box\Phi),
\\
\label{TB2}
\frac{\Box\Phi}{\Phi} &=& - \frac{R}{3}.
\end{eqnarray}

We can now describe the special features of the unimodular KK theory developed above. If we have chosen to make first the dimensional reduction to 4 dimensions and then imposing the unimodular condition, the resulting equations would be the same as the Brans-Dicke unimodular theory \cite{Almeida}, with $\omega = 0$ if $n = 1$, and the information on the Ricci scalar would be lost, as in the usual unimodular theories in four dimensions. This would lead to the Theory A of the figure 2. Instead, we have obtained a new set of equations. In particular, there is now an explicit equation for $R$. Moreover, even in the vacuum case we do not recover the four dimensional scalar-tensor equations in presence of a cosmological constant: instead, a new generalization of the Klein-Gordon equation, with a non-trivial coupling with the geometrical sector. Remark the striking difference of equations (\ref{TB1},\ref{TB2}) with the BD unimodular theory with $\omega = 0$, which is related to the KK theory through the procedure leading to Theory B in figure 2. 

The introduction of matter will lead to richer structure as described below.

\subsection{Introducing matter}

The energy-momentum tensor for a fluid in $n$ dimensions is given by
\begin{eqnarray}
\tilde T^{AB} = (\rho + p)u^A u^B - p g^{AB}.
\end{eqnarray}
We define its traceless part as
\begin{eqnarray}
\tilde\tau_{AB} = \tilde T_{AB} - \frac{1}{D}\tilde g_{AB}\tilde T.
\end{eqnarray}
We use co-moving coordinates such that,
\begin{eqnarray}
u^A = (1,\vec 0).
\end{eqnarray}

We will distinguish the pressure in the external space, denoted by $p_e$, and the pressure in the internal space, denoted by $p_i$.
A direct computation leads to,
\begin{eqnarray}
\tilde \tau_{00} &=& \frac{1}{n + 4}\biggr\{(n + 3)\rho + 3 p_e + n p_i\biggl\},\\
\tilde \tau_{ij} &=& \frac{a^2\delta_{ij}}{n + 4}\biggr\{\rho + (1 + n)p_e - n p_i\biggl\},\\
\tilde \tau_{ab} &=& \frac{\Phi^2\delta_{ab}}{n + 4}\biggr\{\rho - 3p_e + 4p_i\biggl\}.
\end{eqnarray}

The components of the unimodular gravitational tensor, for a flat LFRW metric, are:
\begin{eqnarray}
\tilde E_{00} &=& \frac{1}{n + 4}\biggr\{ - 3(n + 2)\dot H - 3nH^2 - n(n + 2)\frac{\ddot\Phi}{\Phi} + 6nH\frac{\dot\Phi}{\Phi} + n(n  - 1)\frac{\dot\Phi^2}{\Phi^2}\biggl\},\\
\tilde E_{ij} &=& \frac{a^2\delta_{ij}}{n + 4}\biggr\{ (n - 2)\dot H + 3nH^2 - 2n\frac{\ddot\Phi}{\Phi} + n(n - 2)H\frac{\dot\Phi}{\Phi} - n(n  - 1)\frac{\dot\Phi^2}{\Phi^2}\biggl\},\\
\tilde E_{ab} &=& \frac{\Phi^2\delta_{ab}}{n + 4}\biggr\{ - 6(\dot H + 2H^2) - (n - 4)\biggr(\frac{\ddot\Phi}{\Phi} + 3H\frac{\dot\Phi}{\Phi}\biggl) + 4(n  - 1)\frac{\dot\Phi^2}{\Phi^2}\biggl\}.
\end{eqnarray}

The equations of motion are:
\begin{eqnarray}
& &- 3(n + 2)\dot H - 3nH^2 - n(n + 2)\frac{\ddot\Phi}{\Phi} + 6nH\frac{\dot\Phi}{\Phi} + n(n  - 1)\frac{\dot\Phi^2}{\Phi^2} \nonumber\\
& & = 8\pi G\biggr\{(n + 3)\rho + 3 p_e + n p_i\biggl\};\\
& &(n - 2)\dot H + 3nH^2 - 2n\frac{\ddot\Phi}{\Phi} + n(n - 2)H\frac{\dot\Phi}{\Phi} - n(n  - 1)\frac{\dot\Phi^2}{\Phi^2}\nonumber \\ & &= 8\pi G\biggr\{\rho + (1 + n)p_e - n p_i\biggl\},\\
& & - 6(\dot H + 2H^2) - (n - 4)\biggr(\frac{\ddot\Phi}{\Phi} + 3H\frac{\dot\Phi}{\Phi}\biggl) + 4(n  - 1)\frac{\dot\Phi^2}{\Phi^2}\biggl\}\nonumber\\
& &= 8\pi G\biggr\{\rho - 3p_e + 4p_i\biggl\}.
\end{eqnarray}

\subsection{Solving the equations}

First let us consider the vacuum solutions represented by equations (\ref{TB1},\ref{TB2}). In the cosmological set up, this leads to the equations,
\begin{eqnarray}
\dot H + H^2 &=& H\frac{\dot\Phi}{\Phi},\\
\dot H + 3H^2 &=& \frac{\ddot\Phi}{\Phi} + 2H\frac{\dot\Phi}{\Phi}.
\end{eqnarray}
There is a power law solution represented by,
\begin{eqnarray}
a &\propto& t^{1/2},\\
\Phi &\propto& t^{-1/2}.
\end{eqnarray}
The extra dimension contracts as the universe expands. 
Exponential solutions are also possible, but they display an isotropic five-dimensional expansion of all four spatial dimensions. However, it is possible to have a singularity-free solution at least from the point of view of the three dimensional spatial dimension. If fact the equations can be written as
\begin{eqnarray}
\dot H + 2H^2 = k,
\end{eqnarray}
where $k$ is an integration constant. If $k = 0$ we come back to the power law solutions described above. If $k > 0$, on the other hand, we find the solutions,
\begin{eqnarray}
a &\propto& \cosh^{1/2} \sqrt{2k}t,\\
\Phi &\propto& \frac{\sinh \sqrt{2k}t}{\cosh^{1/2} \sqrt{2k}t}.
\end{eqnarray}
This solution display a bounce in the external scale factor and a change of the sign in the field $\Phi$ across the bouncing. 
If $k < 0$, the hyperbolic function are replaced by trigonometric functions and big bang and big crunch singularities appear. Remark that all these solutions imply $R = $ constant, a property of the vacuum case resulting from the application of the Bianchi identities to (\ref{TB1}). This feature will be discussed later. 

The introduction of matter lead to more involved equations. We will discuss some particular, but possibly appealing, cases.

Let us consider now the five-dimensional KK theory ($n = 1$) without pressure in both the external and internal spaces.
The equations of motion under these hypothesis are:
\begin{eqnarray}
- 9\dot H - 3H^2 - 3\frac{\ddot\Phi}{\Phi} + 6H\frac{\dot\Phi}{\Phi}  &=& 32\pi G\rho;\\
-\dot H + 3H^2 - 2\frac{\ddot\Phi}{\Phi} - H\frac{\dot\Phi}{\Phi} &=& 8\pi G\rho;\\
- 6(\dot H + 2H^2) + 3\frac{\ddot\Phi}{\Phi} + 9H\frac{\dot\Phi}{\Phi}
&=& 8\pi G\rho.
\end{eqnarray}

These equation can be combined leading to,
\begin{eqnarray}
- 9\dot H - 3H^2 - 3\frac{\ddot\Phi}{\Phi} + 6H\frac{\dot\Phi}{\Phi}  &=& 32\pi G\rho,\\
\dot H + 3H^2 - \frac{\ddot\Phi}{\Phi} - 2H\frac{\dot\Phi}{\Phi}
&=& 0.
\end{eqnarray}
There are just two independent equations.
The system, even reduced to four dimensions, remains undetermined since we have just two equations for three variables:
$a$, $\Phi$ and $\rho$. 

One particular important case is when the internal space is static. In this case, we can solve the equations obtaining,
\begin{eqnarray}
a \propto t^{1/3}.
\end{eqnarray}
This is equivalent to standard solutions with stiff matter.

The system of equations is remains undetermined even if the pressure in the internal dimension is different from the pressure in 
the external dimension. For example, supposing $p_e = 0$ and $p_i = - \rho$, and considering again a static internal dimension, the scale factor behaves as
\begin{eqnarray}
a \propto t.
\end{eqnarray}
On the other hand, if again the internal dimension is static, but $p_e = - \rho$ and $p_i = 0$, we find a de Sitter phase in the external space:
\begin{eqnarray}
a \propto e^{Ht}, \quad H = \mbox{constant}.
\end{eqnarray}
However, in opposition to the two first cases, the energy density must be negative, and the solution is probably unstable.

One important general feature is that, after reduction to four dimension, the properties of the system of equations are not determined 
anymore by the combination $\rho + p$ as in the isotropic four dimensional unimodular theory \cite{velten}, unless the pressure is the same 
in the internal and external spaces.

Now we compare the above results with the case of the theory in five-dimensions without the unimodular constraint being imposed. 
The equations, after reduction to four dimensions, are:
\begin{eqnarray}
G_{\mu\nu} - \biggr\{\frac{\Phi_{;\mu;\nu}}{\Phi} - g_{\mu\nu}\frac{\Box\Phi}{\Phi}\biggl\}  = 8\pi G \tilde T_{\mu\nu},\\
\frac{\Phi^2}{2}R\delta_{ab} = 8\pi G \tilde T_{ab}.
\end{eqnarray}
For a static internal dimension, the equations reduce to,
\begin{eqnarray}
3H^2 &=& 8\pi G \rho, \\
2\dot H + 3H^2 &=& -8\pi G p_e,\\
3\dot H + 6H^2 &=& 8\pi G p_i.
\end{eqnarray}
When the pressures are absent ($p_e = p_i = 0
$), combination of the two last equations lead to $\dot H = H = 0$, implying $\rho = 0$. The only consistent solution is the Minkowski space-time, 
indicating a static universe. If $p_e = - \rho$, combination of the two first equation leads to the same conclusion, as well as the case $p_i = - \rho$ after combining the first and third equation. Hence, for the three cases considered in the unimodular case, the only solution is the Minkowski space-time, with no matter.

\section{Conclusion}

Unimodular gravity is one of the oldest alternatives to General Relativity theory. It is based on introducing a constraint 
that fixes the determinant of the metric. Few years after unimodular theory has been proposed, the Kaluza-Klein theory has been formulated 
in view of unification of the gravitational and electromagnetic interaction by enlarging the dimension of the space-time to five introducing 
a (spatial) extra dimension besides the known four dimensions. Having combined here both proposals, we constructed a five-dimensional unimodular 
gravity theory. Our main goal was to set up the general structure of the unimodular Kaluza-Klein gravity and its main features after reduction to four dimensions.
As in the more familiar four dimensional unimodular gravity, the unimodular constraint on the determinant of the (now five-dimensional) metric 
restricts the invariance of the theory to the transverse diffeomorphisms. This leads to the emergence of a generalized conservation law for the energy-momentum tensor. 
If, however, the conservation of the five-dimensional energy momentum tensor is imposed, the usual KK five-dimensional equations are recovered with 
a cosmological constant in five dimensions. If the conservation of the five dimensional energy-momentum tensor is not imposed, the KK structure 
can be recovered but with a dynamical cosmological term, similarly to what happens in four dimensions.

Besides setting the general equations, some cosmological solutions in five-dimensions have been obtained.
In vacuum, the unimodular gravity becomes completely equivalent to GR in presence of a cosmological constant. Hence,
the Kasner-type solution of Ref. \cite{CD}, showing spontaneous compactification, is recovered in the KK unimodular gravity. 
In presence of matter the same features of the five dimensional GR theory in presence of a cosmological constant are reproduced in the unimodular context. 
However, unimodular gravity allows a generalized set of conservation laws: relaxing the usual conservation laws, new class of solutions appear. 
In order to implement these solutions an ansatz must be imposed, what is equivalent to introduce a dynamical cosmological term. In particular, 
in the cosmological context, we have shown that solutions with a constant internal dimension are possible even in presence of matter, 
what is not allowed in the usual GR context. 

Another interesting feature is the coupling of the gravity sector to the matter content when the five-dimensional unimodular gravity 
is reduced to four dimensions. As shown, for example, in Ref. \cite{velten}, in the four-dimensional GR unimodular gravity, the equations describing 
the evolution of the universe, in the absence of the usual conservation laws, depend only on the combination $\rho + p$ which may be identified, 
from thermodynamical perspective, with the enthalpy. This remains true in the KK unimodular gravity in a five-dimensional isotropic and homogenous universe 
(which is not realistic, evidently). However, after reducing to four dimensions (or equivalently, by considering a Kasner-type structure in five dimensions), 
this identification is not verified any more, and the pressure acquires a specific rôle, as in the usual GR context.

It is illustrative to set out a complete effective theory in four dimension coupled to matter. If the matter component in the extra dimension is set equal to zero, equations (\ref{TB1},\ref{TB2}) take the following form in presence
of an energy-momentum tensor in four dimensions:
\begin{eqnarray}\label{TB3}
R_{\mu\nu} &=& 8\pi G T_{\mu\nu} + \frac{1}{\Phi}(\Phi_{;\mu;\nu} - g_{\mu\nu}\Box\Phi),
\\
\label{TB4}
\frac{\Box\Phi}{\Phi} &=& 8\pi GT - \frac{R}{3},\\
\label{TB5}
8\pi G {T^{\mu\nu}}_{;\mu} &=& \frac{R^{;\nu}}{2}.
\end{eqnarray}
Hence, in this effective theory in four dimensions, resulting the Kaluza-Klein unimodular theory in five dimensions, the traceless property of the field equations, generally associated to the unimodular theories, disappears, what is not surprising because of construction sketched in the diagram of figure 2. However, equation (\ref{TB5}) reflects the invariance of the theory by a restrict class of diffeomorphism, an important property of the unimodular theory.

In the present analysis we have explored cosmological solutions in the absence of electromagnetic field coming from the KK structure, 
what is required to have isotropy in the three spatial dimensional of the external space. For the KK unimodular gravity the reduction to four dimensions 
preserves the same coupling to the electromagnetic field emerging from the five-dimensional metric. This is related to the traceless character 
of the electromagnetic field. The coupling of the moduli field $\Phi$ associated to the fifth dimension, on the other hand, may possible bring new structures. 
One example, besides the cosmological scenario described here, may be the static, spherically symmetric configurations. Black holes may appear, for example, 
in the Brans-Dicke theory and in the usual KK theory with and without electromagnetic field \cite{k1,k2,clement}, but only in the phantom regime. 
This general feature may change in the KK unimodular gravity. This problem deserves to be studied. For the cosmological solution found here, 
we may also expect some peculiar features at perturbative level. Finally, the effective theory in four dimension, and its coupling to matter emerging 
from KK unimodular, must be developed further since it corresponds to a new implementation of the coupling of the scalar field and gravity after reduction 
to the usual four-dimensional space-time.

\bigskip
\noindent
{\bf Appendix I}

Let us derive the set of general formulas for metrics, connections and curvature in $5$ dimensions, 
with all the $15$ degrees of freedom present.

The calculus in coordinates turns out to be quite complicated, but introducing the non-holonomic local frames 
simplifies the computations considerably.

%The basis of $1$-forms is chosen to be:
%
%\begin{equation}
%\theta^{\mu} = d x^{\mu}, \; \; \; \; \theta^5 = dx^5 + k \; A_{\mu} dx^{\mu},
%\label{thetas}
%\end{equation}
%\textcolor{blue}{to define $k$?}
%

\begin{itemize}

\item Non-holonomic local frame.

The basis of $1$-forms is chosen to be:

\begin{equation}
\theta^{\mu} = d x^{\mu}, \; \; \; \; \theta^5 = dx^5 + k \; A_{\mu} dx^{\mu},
\label{thetas2}
\end{equation}
the constant $k$ must have the dimension of length ($cm$) in order to ensure the uniformity with space variables $x^{\mu}$.

%The $1$-forms:
%\begin{equation}
%{\rm The \; 1-forms:} \; \; \; \theta^{\mu} = d x^{\mu}, \; \; \; \; \theta^5 = dx^5 + k \; A_{\mu} dx^{\mu},
%\label{thetas}
%\end{equation}
The dual vector fields, satisfying $\theta^A ({\cal{D}}_B) = \delta^A_B$:
\begin{equation}
{\cal{D}}_{\mu} = \partial_{\mu} - k \; A_{\mu} \partial_5, \; \; \; \; {\cal{D}}_5 = \partial_5.
\label{CalD2}
\end{equation}
Introducing transition matrices $U^A_B$ and ${\overset{-1}{U^B_C}}$ such that $\theta^A = U^A_B dx^B, \; {\cal{D}}_C = {\overset{-1}{U^D_C}} \partial_D $
we can write:
$$U^{\mu}_{\nu} = \delta^{\mu}_{\nu}, \; \; \; U^{\mu}_5 = 0, \; \; U^5_{\mu} = k A_{\mu}, \; \; U^5_5 = 1;$$
\begin{equation}
{\overset{-1}{U^{\mu}_{\nu}}} = \delta^{\mu}_{\nu}, \; \; {\overset{-1}{U^{5}_{\nu}}} = - k A_{\nu}, \; \; {\overset{-1}{U^{\mu}_{5}}} = 0 
\; \;  {\overset{-1}{U^{5}_{5}}} = 1.
\label{UUinv2}  
\end{equation}
Quite obviously, both determinants are equal to $1$: det(($U^A_B) = 1$ and det (${\overset{-1}{U^B_C}}) = 1.$ 
The metric tensor expressed in the non-holonomic frame is easily deduced from the square of the  $5$-dimensional length element,
taking on the following form:
\begin{equation}
ds^2 = g_{\mu \nu} dx^{\mu} dx^{\nu} - \Phi^2 \left[ dx^5 + k \; A_{\mu} dx^{\mu} \right] \left[ dx^5 + k \; A_{\nu} dx^{\nu} \right]
\label{deessquare2}
\end{equation}
leading to the following $5 \times 5$ matrix representation:
\begin{equation}
{\tilde{g}}^{AB} = \begin{pmatrix} g_{\mu \nu} + k^2 \Phi^2 A_{\mu} A_{\nu} & - k \Phi^2 A_{\nu} \cr - k \Phi^2 A_{\mu} & - \Phi^2 \end{pmatrix} 
\label{gABfive2}
\end{equation}
The inverse matrix becomes then:
\begin{equation}
{\tilde{g}}^{BC} = \begin{pmatrix} g^{\nu \lambda}  &  k A_{\lambda} \cr k  A_{\nu} & - \Phi^{-2} + k^2 A^{\nu} A^{\lambda} \end{pmatrix} 
\label{invgABfive2}
\end{equation}
One easily checks that 
$${\tilde{g}}_{AB} {\tilde{g}}^{BC} = \delta^A_C.$$
It is also obvious that due to the fact that the transition matrix $U^A_B$ is unimodular, the determinant of the metric ${\tilde{g}}_{AB}$
is the same as that of the metric $g_{AB}$, i.e. equal to $\Phi^2$. 
The simplest and most elegant way to evaluate the connection coefficients and the components of the Riemann tensor   
is to use the non-holonomic frame $\theta^A$ and its dual basis of derivations (vector fields) ${\cal{D}}_B, \; \; A, B = 1,2...5.$

We need to know the commutators of non-holonomic derivations. We have:
\begin{equation}
\left[ {\cal{D}}_A, {\cal{D}}_B \right] = C_{AB}^E \; {\cal{D}}_E,
\label{commDADB}
\end{equation}
where
\begin{equation} 
C^{5}_{\mu \nu} = C_{\mu \nu 5} = - k \; F_{\mu \nu} = -k \; (\partial_{\mu} A_{\nu} - \partial_{\nu} A_{\mu} ).
\end{equation}
We have then the connection coefficients in the non-holonomic basis:
$${\hat{\Gamma}}^{C}_{AB} = \frac{1}{2} {\hat{g}}^{CE} \left[ {\cal{D}}_A g_{BE} + {\cal{D}}_B g_{AE} - {\cal{D}}_E g_{AB} \right] + $$
\begin{equation}
{\hat{g}}^{CE} \left[ C_{EAB} + C_{EBA} - C_{BAE} \right]
\end{equation}
where ``hat'' refers to the components with respect to the anholonomic frame.

\item Connection.

We have then the connection coefficients in the non-holonomic basis:
$${\hat{\Gamma}}^{C}_{AB} = \frac{1}{2} {\hat{g}}^{CE} \left[ {\cal{D}}_A g_{BE} + {\cal{D}}_B g_{AE} - {\cal{D}}_E g_{AB} \right] + $$
\begin{equation}
{\hat{g}}^{CE} \left[ C_{EAB} + C_{EBA} - C_{BAE} \right]
\end{equation}
where the ``hat'' refers to the components with respect to the anholonomic frame.

The only non vanishing connection coefficients are then the following:
\begin{equation}
{\hat{\Gamma}}^{\mu}_{\nu \lambda} = \Gamma^{\mu}_{\nu \lambda}, \; \; {\hat{\Gamma}}^{\mu}_{\nu 5} = {\hat{\Gamma}}^{\mu}_{5 \nu} = - \frac{1}{2} k F^{\mu}_{\; \; \nu},
\; \;  {\hat{\Gamma}}^{5}_{\nu \lambda} = - {\hat{\Gamma}}^{5}_{\lambda \nu} = \frac{1}{2} k F_{\lambda \nu},
\label{Gammahat}
\end{equation}

\item Riemannian curvature in $5D$.

The Riemann tensor expressed in a non-holonomic frame is:

\begin{equation}
{\hat{R}}_{ABD}^C = {\cal{D}}_A {\hat{\Gamma}}^C_{BD} - {\cal{D}}_B {\hat{\Gamma}}^C_{AD} + {\hat{\Gamma}}^C_{AF} {\hat{\Gamma}}^F_{BD}
- {\hat{\Gamma}}^C_{BF} {\hat{\Gamma}}^F_{AD} - C^F_{AB} {\hat{\Gamma}}^C_{FD} 
\label{Riemanntensor}
\end{equation}
The Ricci tensor and the curvature scalar in $5$ dimensions are calculated as usual,
\begin{equation}
{\hat{R}}_{AD} = {\hat{R}}_{ACD}^C, \; \; \; \; \; {\hat{R}} = {\hat{g}}^{AB} {\hat{R}}_{AB}.
\label{RicciR}
\end{equation}

\end{itemize}

\bigskip
\noindent
{\bf Acknowledgement:} JCF thanks CNPq and FAPES for partial financial support. RK would like to acknowledge Philip D. Mannheim's valuable comments 
and suggestions }

\end{document}